**Migration volume for polaron dielectric relaxation in disordered materials**

A.N. Papathanassiou [a)], I. Sakellis and J. Grammatikakis

University of Athens, Department of Physics, Section of Solid State Physics,
Panepistimiopolis, GR 157 84 Zografos, Athens, Greece

[a)] Author to whom correspondence should be addressed; E-mail address:
antpapa@phys.uoa.gr

**Abstract**

A theoretical study of the influence of pressure on the dielectric relaxation related with polaron tunneling and phonon assisted hopping in disordered solids is developed. The sign and absolute value of the migration volume, which is obtained by employing the present formulation, evidence about the nature of the relaxation. As a paradigm, positive and negative values of migration volume are evaluated by analyzing recently published dielectric loss measurements under pressure in semiconducting polypyrrole. A straightforward relation between the value of the migration volume and the nature of short-range polaron flow and the size of polaron distortion is revealed.

Polaron hopping in disordered matter under the action of a harmonic electric field results in dielectric relaxation. Ac conductivity vs frequency is dispersive and a dielectric loss peak is detected within the complex impedance representation [1-6]. The relaxation time $\tau$, which governs the relaxation process and determines the position of the maximum of the loss peak in the frequency domain through the resonance condition $\omega_m\tau=1$ ($\omega_m$ is the angular frequency where the loss peak exhibits a maximum), was revisited by Mott and Davis [1] by combining quantum mechanical tunneling and phonon assisted hopping. The *temperature* evolution of the relaxation time has been used to obtain the value of the "activation energy" from dielectric measurements at various temperatures [2-5]. On the contrary, a detailed theoretical interpretation of the *pressure* evolution of the relaxation time is missing. The theoretical interpretation of dielectric measurements under pressure in ionic crystals and organic solids, where non-electronic relaxation occurs, was actually established in the past [7-14]. In the present work, we develop analytical formulas that permit the quantitative interpretation of polaron (dielectric) relaxation under pressure in *disordered media*. As a paradigm, dielectric measurements vs pressure in conducting polypyrrole [6] are analyzed. The value of the migration volume for relaxation is obtained. Conducting polymer networks are regarded as disordered solids, where the variable range hopping model applies [2, 3, 15] and is employed to analyze the ac conductivity dispersion and the dielectric loss peaks.

An electric dipole of a dielectric relaxes classically by successive motion of its constituting electric charges along neighboring equilibrium sites by overcoming the potential barrier separating them. The relaxation time $\tau$ is [16]:

$$\tau = \nu^{-1}\exp\left(\frac{g^m}{kT}\right) \qquad (1)$$

where $\nu$ is the attempt frequency of the electric charge entity towards the potential barrier, $g^m$ is the migration Gibbs energy, k is the Boltzmann's constant and T is the absolute temperature. A constant related with the degrees of freedom of the dipole, can be included in the right hand side of Eq. (1). All thermodynamic quantities appearing in this paper are related with relaxation or, alternatively, with the localized motion of the electric charge entities that form the relaxing 'dipoles'.

Relaxation in disordered media occurs as polarons (or electrons) hop between localized sites at a distance R from each other, separated by a potential barrier. The relaxation time incorporates both tunneling and phonon-assisted hopping [1]:

$$\tau = \nu^{-1} \exp(2\alpha R) \exp(E/kT) \qquad (2)$$

where $\alpha$ is the inverse localization length of the wave-function and E is the activation energy required to surmount the barrier separating neighboring sites. $\nu^{-1} \exp(E/kT)$ is identical to the right hand side of Eq. (1), implying that the "activation energy" coincides with $g^m = h^m - Ts^m$, where $h^m$ and $s^m$ denote the migration enthalpy and entropy, respectively. By setting $\tau_0 \equiv \nu^{-1} \exp(-s^m/k) \exp(2\alpha R)$, which is slightly dependent on temperature, eq. (2) takes the well-known Arrhenius form $\tau = \tau_0 \exp(h^m/kT)$. The Mott and Davis relaxation time (Eq.(2)) can be re-written as follows:

$$\tau = \nu^{-1} \exp(2\alpha R) \exp(g^m/kT) \qquad (3)$$

Differentiating the (natural) logarithm of the last equation with respect to pressure at constant temperature, we get:

$$\left(\frac{\partial \ln \tau}{\partial P}\right)_T = \left(-\frac{\partial \ln \nu}{\partial P}\right)_T + 2R\left(\frac{\partial \alpha}{\partial P}\right)_T + 2\alpha\left(\frac{\partial R}{\partial P}\right)_T + \frac{1}{kT}\left(\frac{\partial g^m}{\partial P}\right)_T \qquad (4)$$

The first term of the right-hand side is: $\left(-\dfrac{\partial \ln \nu}{\partial P}\right)_T = \left(-\dfrac{\partial \ln \nu}{\partial \ln V}\right)_T \left(\dfrac{\partial \ln V}{\partial P}\right)_T = -\gamma\chi_T$, where $\gamma \equiv -\left(\dfrac{\partial \ln \nu}{\partial \ln V}\right)_T$ is the Grüneisen constant (V is the volume) and

$\chi_T \equiv -\left(\dfrac{\partial \ln V}{\partial P}\right)_T$ is the isothermal compressibility. The modification of R with

pressure is $\left(\dfrac{\partial R}{\partial P}\right)_T = \left(R\dfrac{\partial \ln R}{\partial P}\right)_T = \left(R\dfrac{1}{3}\dfrac{\partial \ln R^3}{\partial P}\right)_T = \left(R\dfrac{1}{3}\dfrac{\partial \ln V}{\partial P}\right)_T = -\dfrac{1}{3}R\chi_T$.

The latter formula is based on the assumption that the medium is isotropic. The last term of Eq. (4) is related with the migration volume related for polaron relaxation $\upsilon^m \equiv \left(\partial g^m / \partial P\right)_T$ [8-12], which is defined as $\upsilon^m = V_e - V_g$, where $V_e$ and $V_g$ are the volumes of the specimen when the relaxing charge is in its excited and ground state, respectively [8, 13].

Lundin *et al* [17] suggested that the wave-function inverse localization length increases linearly upon pressure $\alpha^{-1} = \alpha^{-1}(P=0)\left[1 + \chi_T P\right]$. Here, we use the more general form $\alpha^{-1} = \alpha^{-1}(P=0)\exp\left(\chi_T P\right)$, which reduces to the aforementioned linear relation, since $e^{\chi_T P} \approx 1 + \chi_T P$, for $\chi_T P \ll 1$ (which is valid in the pressure range dielectric measurements are usually performed) and $\chi_T$ and assuming that $\chi_T$ changes negligibly in a narrow pressure range. We therefore have:

$$\left(\frac{\partial \alpha}{\partial P}\right)_T = -\alpha^2 \left(\frac{\partial \alpha^{-1}}{\partial P}\right)_T = -\alpha \chi_T \qquad (5)$$

Thus, Eq. (4) reduces to:

$$\upsilon^m = kT\left(\frac{\partial \ln \tau}{\partial P}\right)_T + (\gamma - \frac{4}{3}\alpha R)\chi_T \qquad (6)$$

provided that *the inverse localization length increases linearly with pressure*.

On the other hand, Maddison and Tansley [18] asserted that $\alpha$ is substantially pressure independent in conducting polypyrrole, because it describes a bound-state-like wave function, an eigenstate of the rigid monomer benzole ring, and is therefore unlikely to be modified significantly by pressure. Taking $\left(\partial \alpha / \partial P\right)_T = 0$, Eq. (4) reduces to:

$$\upsilon^m = kT\left(\frac{\partial \ln \tau}{\partial P}\right)_T + (\gamma + \frac{2}{3}\alpha R)\chi_T \qquad (7)$$

We stress that the latter is valid under the restriction that *the wave function inverse localization length is insensitive to pressure modification*.

In conducting polypyrrole, which is a disordered organic semiconductor, a broad dielectric loss peak recorded at zero pressure, splits into a couple of distinct components on pressurizing [6] (Figure 1). One constituent (mechanism I) is insensitive to the increase of pressure, while another (mechanism II) shifts gradually towards higher frequency on increasing the pressure. Relaxation I is attributed to intra-chain (intra-cluster) charge flow and relaxation II to inter-chain (inter-cluster) hopping through the void space separating neighboring chains or grains [6]. An increase of hydrostatic pressure affects mainly the inter-chain (or, most likely, the inter-grain) space and, subsequently, the inter-chain (or inter-cluster) hopping. Intra-chain (or intra-grain) conductivity effects are much weaker since it is more energetically favorable to achieve conformational re-arrangement as the sample volume is suppressed, than reducing the length of individual chain (and the intra-chain length, respectively) [6].

The (natural) logarithm of $\tau$ vs pressure is best described by a linear law rather than a second order polynomial one [7] (Figure 1). The values of $\upsilon^m$ for relaxation in conducting polypyrrole obtained from the pressure variation of the relaxation time at room temperature, through Eqs. (6) and (7), respectively, are shown in Table I. A Grüneisen parameter $\gamma \approx 4.0 \pm 0.1$ [19], typical for polymeric solids was employed (more recent progress can be found in Refs. 20 and 21). This value is larger than the value 1.5-1.7 of metals, ionic and covalent solids share. 2aR was set equal to unity [1]. A couple of zero pressure isothermal compressibility values [17,22] were used.

$\upsilon^m$ is positive for relaxation mechanisms I and negative for mechanism II, respectively (Table I). The physical content of the activation volume for protonic motion was given initially by Fontanella *et al* [13], as the volume change of the material induced when the transferring charge undergoes a transition from a 'ground' state to an 'excited' one. The *positive* value of $\upsilon^m$ implies that relaxation I induces an *outwards* relaxation of the solid for intra-chain transport (mechanism I). Inter-chain hopping occurs by the passage of the polaron through the inter-separating void space ('excited' state). At this instance, the volume distortion is less than that induced when the polaron is located at the chain ('ground' state), and the resulting $\upsilon^{act}$ is *negative* (Relaxation II), indicating an *inward* relaxation of the polymer network on the 'excited' state of inter-chain hopping [6]. The polaron radius in polypyrrole is $r_P \approx 1.2 \overset{\circ}{A}$ [23], which yields a volume $\frac{4}{3}\pi r_p^3 \approx 7.2 \overset{\circ}{A}^3$ The value of the $\upsilon^m$ for relaxation II (Table I) evidences for a volume contraction comparable with the polaron volume: e.g., the absolute value of $\upsilon^m = -8.7 \overset{\circ}{A}^3 \pm 10\%$ is practically equal to the polaron volume. At the instance when the polaron crosses the void space ("excited" state), the macroscopic distortion is nearly null. So, the difference between the macroscopic volume when the charge crosses the void space ('excited' state) and the one when the polaron settles at a polymer chain should be negative and nearly equal to the polaron volume. As we saw, this picture is quantitatively justified by comparing the $\upsilon^m$ for relaxation II with the polaron volume.

In summary, the pressure derivative of the dielectric relaxation time in disordered media (where tunneling and phonon-assisted hopping occur) was expressed in terms of the Grüneisen parameter, the isothermal compressibility and the migration volume. The migration volume for relaxation can be obtained from dielectric measurements under pressure. As a paradigm, the present methodology was applied to analyze recently published results in conducting polypyrrole, where two different relaxation modes with entirely converging behavior are traced. Positive and negative values of the migration volume are evaluated, corresponding to intra and iter-chain charge flow, respectively. A correlation with the polaron radius seems to justify the microscopic description of conduction modes.

## Acknowledgements

The authors are grateful to Prof. P.A. Varotsos for his potential comments and helpful recommendations.

**Table I.** Evaluation of $\upsilon^m$ for two relaxation mechanisms in conducting polypyrrole from the pressure variation of $\tau$ at room temperature (292 K), taking into account the correction terms appearing in Eqs. (6) and (7) respectively. Zero pressure isothermal compressibility values are from two sources: [b] $\chi_T(P=0)= 0.061$ GPa$^{-1}$ from Ref. 17 and [c] $\chi_T(P=0)= 0.08$ GPa$^{-1}$ from Ref. 22. Details are given in the text.

| Dielectric loss mechanism | $\left(\dfrac{\partial \ln \tau}{\partial P}\right)_T$ (GPa$^{-1}$) | Eq. (6) (pressure dependent $\alpha$) | | | Eq. (7) (constant $\alpha$) | | |
|---|---|---|---|---|---|---|---|
| | | $(\gamma-\frac{4}{3}\alpha R)\chi_T(P=0)$ (GPa$^{-1}$) | $\upsilon^m/kT$ (GPa$^{-1}$) | $\upsilon^m$ $\left(\overset{\circ}{A}^{3}\right)$ | $(\gamma+\frac{2}{3}\alpha R)\chi_T(P=0)$ (GPa$^{-1}$) | $\upsilon^m/kT$ (GPa$^{-1}$) | $\upsilon^m$ $\left(\overset{\circ}{A}^{3}\right)$ |
| I | $\approx 0.0 \pm 0.1$ | 0.264 [b] | 0.264 | 1.1 | 0.203 [b] | 0.203 | 0.8 |
| | | 0.346 [c] | 0.346 | 1.4 | 0.266 [c] | 0.266 | 1.1 |
| II | $-2.5 \pm 0.2$ | 0.264 [b] | -2.236 | -9.0 | 0.203 [b] | -2.297 | -9.3 |
| | | 0.346 [c] | -2.154 | -8.7 | 0.266 [c] | -2.234 | -9.0 |

**Figure Caption**

**Figure 1:**     The natural logarithm of the relaxation time of dielectric loss mechanisms I (squares) and II (circles), respectively, in conducting polypyrrole. Straight lines best fit the data points. Inset: The imaginary part of the dielectric permittivity $\varepsilon''$ (after subtraction of the dc component) as a function of frequency at room temperature for two different pressures: (a): ambient pressure and (b): 0.30 GPa.

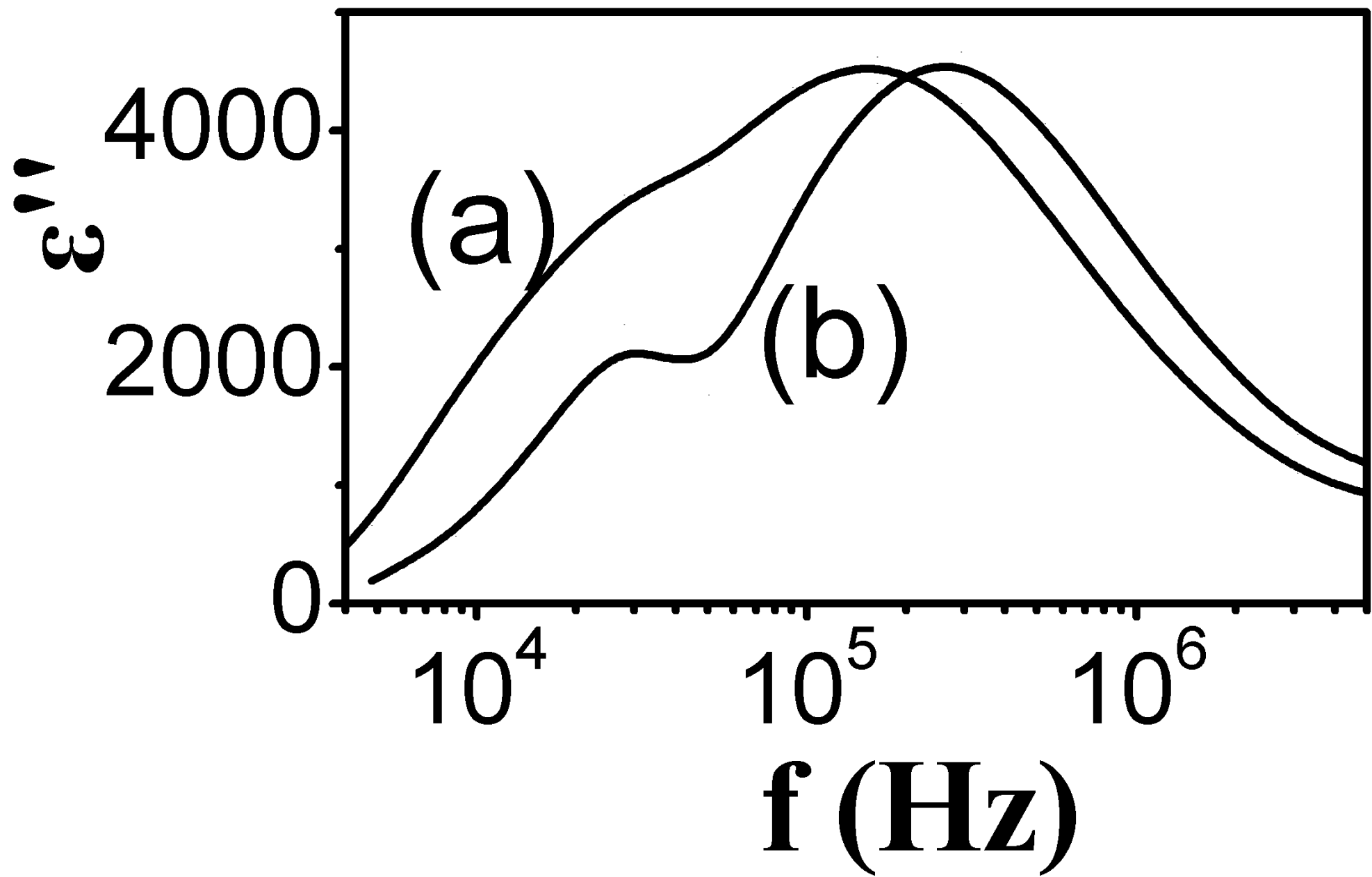